\newcommand{\LHC}{\mbox{LHC}}
\newcommand{\LEP}{\mbox{LEP2}}
\newcommand{\Pythia}{\mbox{PYTHIA}}
\newcommand{\Herwig}{\mbox{HERWIG}}
\newcommand{\MCatNLO}{\mbox{MC@NLO}}
\newcommand{\Jimmy}{\mbox{JIMMY}}
\newcommand{\BHO}{\mbox{BHO}}
\newcommand{\SM}{Standard Model}
\newcommand{\MC}{Monte Carlo}

\newcommand{\beq}{\begin{equation}}
\newcommand{\eeq}{\end{equation}}
\newcommand{\bea}{\begin{eqnarray}}
\newcommand{\eea}{\end{eqnarray}}

\newcommand{\twsq}{\mbox{$\tan^2\theta_w$}}

\newcommand{\epem}{\mbox{$\mathrm{e^+e^-}$}}

\newcommand{\WW}{\mbox{$\mathrm{W^+W^-}$}}
\newcommand{\WZ}{\mbox{$\Wpm\mathrm{Z}$}}
\newcommand{\WpZ}{\mbox{$\Wp\mathrm{Z}$}}
\newcommand{\WmZ}{\mbox{$\Wm\mathrm{Z}$}}
\newcommand{\ZZ}{\mbox{$\mathrm{ZZ}$}}
\newcommand{\Zg}{\mbox{$\mathrm{Z}\gamma$}}
\newcommand{\gaga}{\mbox{$\gamma\gamma$}}
\newcommand{\Wg}{\mbox{$\Wpm\gamma$}}
\newcommand{\Wp}{\mbox{$\mathrm{W^+}$}}
\newcommand{\Wpg}{\mbox{$\Wp\gamma$}}
\newcommand{\Wm}{\mbox{$\mathrm{W^-}$}}
\newcommand{\Wmg}{\mbox{$\Wm\gamma$}}
\newcommand{\Wpm}{\mbox{$\mathrm{W^\pm}$}}

\newcommand{\qbar}{\mbox{$\mathrm{\overline{q}}$}}

\newcommand{\Phist}{\mbox{$\phi^{*} $}}
\newcommand{\Thest}{\mbox{$\theta^{*} $}}

\newcommand{\pt}{\mbox{$p_T$}}

\newcommand{\lgg}{\mbox{$\lambda_\gamma$}}

\newcommand{\lz}{\mbox{$\lambda_{\mathrm{z}}$}}
\newcommand{\dgz}{\mbox{$\Delta g_1^{\mathrm{z}}$}}

\newcommand{\dkg}{\mbox{$\Delta \kappa_\gamma$}}
\newcommand{\dkz}{\mbox{$\Delta \kappa_{\mathrm{z}}$}}
\documentstyle[epsfig,a4p,cite,12pt]{article}

\begin{document}
\bibliographystyle{plain}
\begin{titlepage}
\bigskip\bigskip\bigskip\bigskip\bigskip
\begin{center}
 {\LARGE\bf \boldmath 
Weighting Di-Boson \MC~Events in Hadron Colliders}
\end{center}
\bigskip\bigskip
\begin{center}
{\large Gideon Bella}\footnote{e-mail: {\tt bella@atlas2.tau.ac.il}} 
\end{center}
\bigskip
\begin{center}
{\large Raymond and Beverly Sackler School
  of Physics and Astronomy, Tel Aviv University, Tel Aviv, Israel}
\end{center}
\bigskip\bigskip
\begin{center}
{\Large  Abstract}
\end{center}
A detailed study of the di-boson \MC~programs \Pythia, \MCatNLO~and
the program of Baur, Han and Ohnemus (\BHO) is performed. None of
these programs cover all aspects of di-boson production. 
The \BHO~code is used to produce event weights emulating anomalous 
triple gauge couplings in \Pythia~and \MCatNLO~events. In the same way,
boson spin information which is missing for most di-boson channels in
\MCatNLO~can be introduced as well. This weighting code can be used
to study systematic effects related to various aspects of the \MC~generators, 
e.g. parton distribution functions. A detailed study comparing
distributions of event samples generated with these three generators shows a
nice agreement for events without jets. Some differences between the
three samples are observed for events with jets. Most of these
differences can be attributed to the different ways of jet production in
the three programs. 

{\bf Keywords:} Hadron-Hadron Scattering, Standard Model, Spin and 
Polarization Effects

\end{titlepage}


\section{Introduction}
 \label{sec:intro}

 Di-boson production is one of the most interesting processes to test
the \SM. At \LEP~\epem collider~\cite{LEP2}, \WW~production has been
used to investigate the properties of the W-boson, in particular its
mass and decay branching fractions, and to measure the
W-polarization. In addition, cross sections for
\WW, \ZZ~and \Zg~production have been measured and these processes
have been used to look for anomalous Triple Gauge Couplings (TGC's).

As far as hadron colliders are concerned, the measurement of the
W-boson properties is based on single W
events where the production rate is by far higher than for di-bosons.
Di-boson events are useful for cross-section measurements, TGC and 
polarization studies. Both Tevatron experiments, CDF~\cite{CDF} and
D\O\ \cite{D0} have used their
Run II data of p$\bar{\rm p}$ collisions at center-of-mass energy of
1.96 GeV to measure cross sections for \WW, \WZ, \ZZ, \Wg~and \Zg~
production and to set limits on the anomalous TGC's.
The \LHC~sensitivity for all the anomalous TGC's is expected to improve
with respect to the Tevatron, and for most anomalous couplings, also
with respect to \LEP, due to the higher energies and larger event
samples at the LHC. The TGC analysis is based on
comparing measured and expected distributions of kinematic
observables, and therefore, should
rely on the best possible modelling of the di-boson \MC~production.
Moreover, di-boson events
constitute an important background source for new particle searches,
most notably the Higgs boson which, if massive enough, decays mainly
into \WW~and \ZZ~final states. Consequently, the new particle discovery 
reach is affected by the quality of the \MC~generators of the di-boson
background channels.

\begin{figure}[htbp]
\begin{center}
\epsfig{file=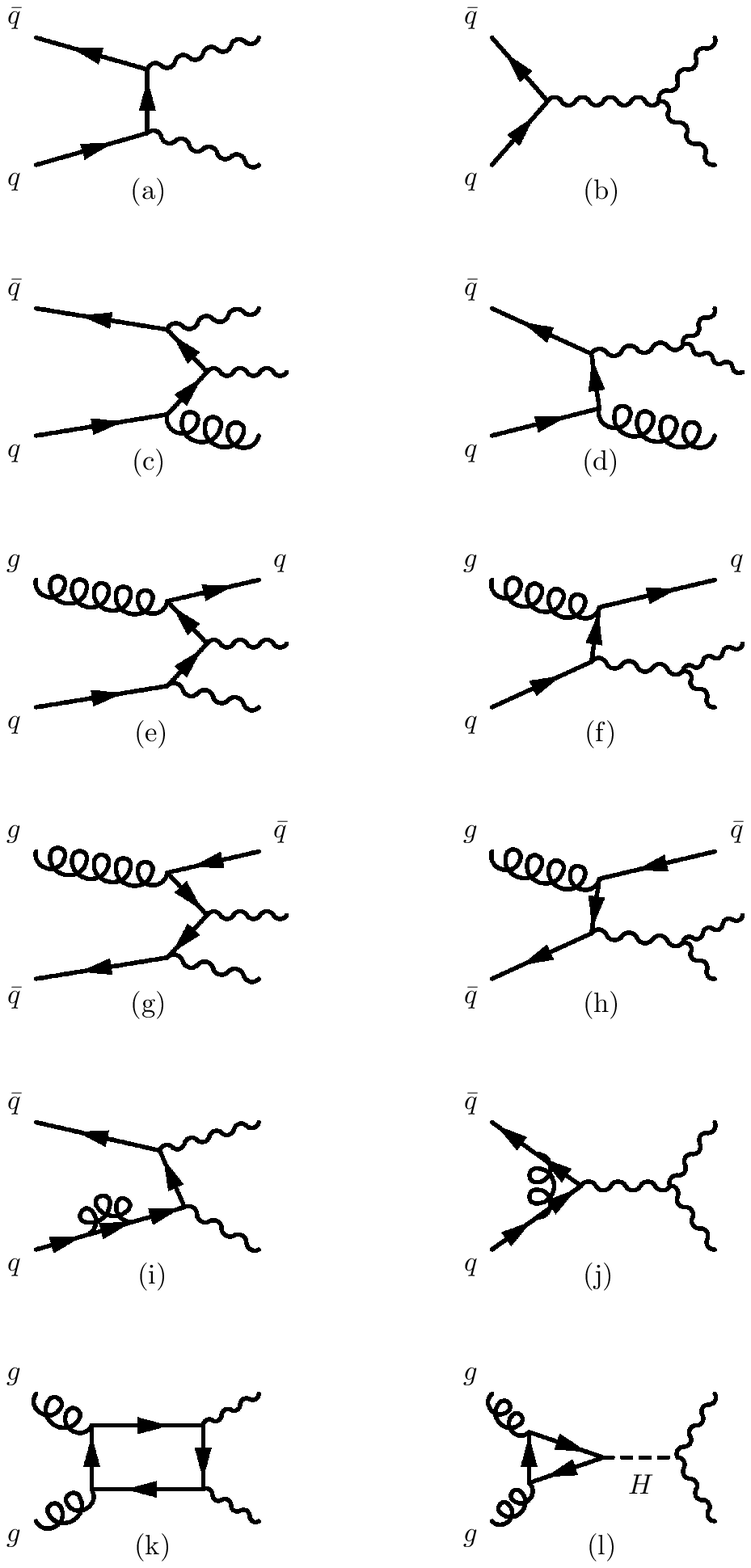,bbllx=165pt,bblly=165pt,bburx=450pt,bbury=735pt,
     clip=}
\caption{\sl
Feynman diagrams contributing to di-boson production at hadron colliders
}
\label{fig:feyn}
\end{center}
\end{figure}

Figure~\ref{fig:feyn} shows many of the Feynman diagrams
contributing to di-boson
production at hadron colliders w/o outgoing partons. Diagrams (a) and
(b) correspond to the 0th (leading) order (LO) in QCD, namely the
Born cross-section without outgoing parton. Diagrams (c) and (d) 
are similar to (a) and (b), with the additional emission of one gluon,
thus being 1st (next-to-leading) order (NLO).
Since gluons are very abundant in the interacting protons,
particularly at the high LHC energy, diagrams such as (e)-(h) should
be included as well. 
One loop diagrams like (i) and (j) are 2nd order in QCD, but they
interfere with the LO diagrams (a) and (b). This interference is then
1st order in QCD and should be included in any NLO calculation. 

Consequently, in LO di-boson \MC~generators, the result of the hard
interaction between the incoming partons is the di-boson system,
whereas NLO generators produce two types of events. The first type
includes the di-boson system only, and is due to the LO diagrams (a)
and (b), and their interference with the loop diagrams (i) and (j). 
One also adds the contribution of
diagrams like (c) and (d), where the outgoing parton is
either soft or collinear. This is needed because there are infra-red 
divergences both for real and virtual partons which cancel each other
when added together. The soft parton is not observed at the final
state. Since the interference term can be negative, the cross section
for many of these events is negative and they are associated with 
negative weights.\footnote{
In most cases, events are generated with some approximate 
distribution, and obtain a weight being equal to the ratio between the
exact cross section value and the approximated expression used for
generation. An unweighted event sample can be obtained, only for the
case where all weights are positive, using a 
random number generator to reject part of the generated events.}
The second type of events produced by the NLO generators is due to 
diagrams (c)-(h) with hard parton which is observed in the final 
state along with the two bosons. All these events have positive cross 
section and are generated with positive weights. 

The gluon-gluon fusion diagrams (k) and (l), which 
can produce only di-boson final states with zero charge 
(e.g. \WW, \ZZ), describe a separate process and do not interfere with the
LO diagrams, thus contributing only to the next-to-next leading order 
(NNLO) in QCD. This is why those diagrams are missing from 
the NLO calculations but nevertheless, their contribution is not
negligible~\cite{ggWW}, again due to the high abundance of gluons in 
the high energy interacting protons. Recently, the \MC~programs gg2WW
and gg2ZZ to generate \WW~and \ZZ~final states have become 
available~\cite{MCgg}. These processes will not be 
further discussed in this note.

Several di-boson \MC~generators exist. Each one of them includes some 
important aspects of the di-boson production, but, as 
will be described
in the next section, none of them cover all the important aspects.
In this report, we suggest to combine the advantages of two
\MC~programs, using one of the programs to generate the events and the
second program to calculate a correction weight for each event in order
to complement the physics aspects missing in the first program.
This event weighting method will be described in the third section, 
and demonstrated with simulated data. The resulting distributions of
kinematic observables of the different \MC~generators will be compared
and the differences will be discussed. 
The technical details of using our weighing program will be described
in the Appendix.  


\section{Monte Carlo Generators}
 \label{sec:mcgen}

We examine three \MC~generator programs, \Pythia6.4, \BHO~and 
\MCatNLO3.2.

The \Pythia~program~\cite{PYTHIA} generates all possible di-boson
pairs, \WW, \WZ, \Wg, \ZZ, \Zg~and \gaga. However, NLO effects are not
included, and there is no implementation of anomalous couplings.

\begin{figure}[htbp]
\begin{center}
\epsfig{file=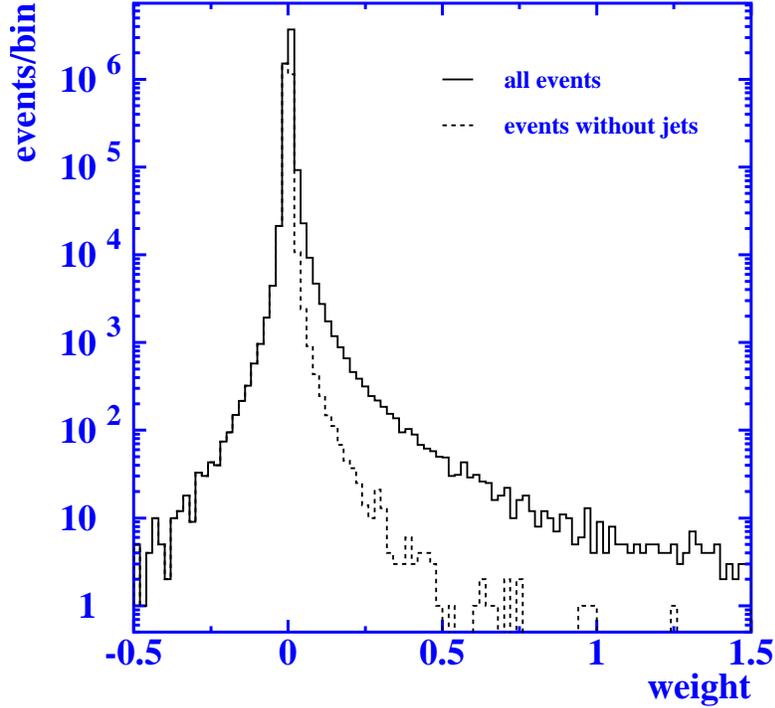,bbllx=18pt,bblly=148pt,bburx=538pt,bbury=630pt,
     width=0.6\textwidth,clip=}
\caption{\sl
Distribution of weights for \WW~events generated for the LHC by the
BHO program
}
\label{fig:weight}
\end{center}
\end{figure}

The program by Baur, Han and Ohnemus (\BHO)~\cite{BHO} generates 
\WW, \WZ, \Wg~and \Zg~events. Charged current WW$\gamma$ (\dkg, \lgg) 
and WWZ (\dkz, \lz, \dgz) anomalous couplings~\cite{HAGIWARA} are 
implemented for \WW, \WZ~and \Wg~production. For \Zg~production, there is an
implementation of neutral current anomalous couplings which are
missing in the standard model, Z$\gamma\gamma^*$ and 
Z$\gamma$Z$^*$($h_i^\gamma$, $h_i^Z$, $i$=1,2,3,4)\footnote{
These are couplings of the intermediate state off-shell photon or Z,
denoted here by $\gamma^*$,Z$^*$, to the outgoing on-shell photon and Z.}
~\cite{HAGIWARA}.
The outgoing gauge bosons are produced with their nominal mass values
without width. The events are weighted and the weight distribution for
\WW~events at the LHC is
shown in Fig.~\ref{fig:weight}. This distribution is very broad,
with a large number of negative weights. All these negative weights
are associated with events of pure
di-boson final state without any outgoing partons. Therefore, it is not
possible to produce a sample of unweighted events. Moreover, having
most of the events produced with very small weights is a serious
problem for the LHC where a significant amount of CPU time is needed 
for the detector simulation of each event. Only the hard interaction 
between the incoming partons
is simulated using the Matrix Element (ME) of the process, generating
the decay products of the gauge bosons and possibly
one outgoing parton, henceforth referred as ME parton or ME jet.
However, the underlying event is ignored, and there is neither 
parton showering, nor hadronization, which is a serious disadvantage.  
Including the simulation of these missing parts, could 
be done by \Pythia~or \Herwig~\cite{HERWIG}, using e.g. the Les Houches 
Accord~\cite{LHA}. However, it would introduce double counting of the 
ME parton, and the other partons produced in the parton shower 
simulation, henceforth referred as SH partons or SH
jets. Consequently, BHO generated events cannot be used as input for
full detector simulation.

The program \MCatNLO~\cite{MC@NLO}~generates \WW, \WZ, \ZZ~and some 
other non di-boson processes. The generation is done in two steps. 
The first step, called NLO, generates only the hard interaction, 
producing the di-bosons and, for part
of the events, the outgoing ME parton, just like the \BHO~generator.
The generated events are written to a file. The second step, called
MC, reads the generated events from the file and process them with 
\Herwig~to simulate the underlying events, the parton shower and
hadronization processes. Special care is taken in the parton shower process 
of \Herwig~to avoid double counting of ME and SH partons. Typically, 85\%
of the events are generated with weight of 1 and the rest with weight
of -1. Multiple interactions between the incoming partons, included in
\Pythia, are not part of \Herwig, but can be still included by running
\Herwig~in the MC step together with the \Jimmy~program~\cite{JIMMY}.
As in BHO, the gauge bosons are produced with their nominal mass
without width. Unfortunately, anomalous couplings are not implemented.
Apart from W-pair production where both W's decay into leptons, the
decay of the gauge bosons is done in the second step, which does not 
have the information on the boson helicity states. Therefore, the
gauge bosons decay isotropically. This is a problem mainly for
analysis aiming at polarization measurements.

Table~\ref{tab:properties} summarizes the characteristics of the
three different \MC~programs described above.  

\begin{table}[thbp]
 \begin{center}
 \begin{tabular}{|l|c|c|c|} \hline
 Generator & \Pythia & BHO & \MCatNLO+\Jimmy \\ \hline
 Processes &  all & \WW, \WZ, \Wg, \Zg & \WW, \WZ, \ZZ \\ \hline
 NLO & $\times$ & $\surd$ & $\surd$ \\ \hline
 Boson width & $\surd$ & $\times$ & $\times$ \\ \hline
 Spin information & $\surd$ & $\surd$ &  $\times$ \\ \hline
 Anomalous TGC &  $\times$ & $\surd$ &  $\times$ \\ \hline
 Fragmentation, hadronization, & $\surd$ &  $\times$ & $\surd$ \\
 underlying event & & & \\ \hline
 \end{tabular}
 \caption{Properties of the different \MC~programs} 
 \label{tab:properties}
 \end{center}
\end{table}

\begin{table}[thbp]
 \begin{center}
 \begin{tabular}{|l|l|c|cc|c|} \hline
  & Generator & \Pythia & \multicolumn{2}{|c|}{\BHO} & \MCatNLO+\Jimmy
  \\ \hline \hline
  &  & LO & LO & NLO & NLO \\
  & Total cross section [pb] & 3.510 & 3.774 & 4.978 & 5.211 \\ \cline{2-6}
\WW  & Fraction of events with jets & 0.268 & \multicolumn{2}{|c|}{0.478} &  
 0.311 \\ \cline{2-6} 
  & Fraction of g,q,\qbar~jets & 1., 0., 0. & 
  \multicolumn{2}{|c|}{0.37, 0.47, 0.16} & 0.58, 0.31, 0.11 \\ \hline \hline
  &  & LO & LO & NLO & NLO \\
  & Total cross section [pb] & 0.257 & 0.273 & 0.417 & 0.432 \\ \cline{2-6}
\WpZ  & Fraction of events with jets & 0.294 & \multicolumn{2}{|c|}{0.347} &  
 0.474 \\ \cline{2-6} 
  & Fraction of g,q,\qbar~jets & 1., 0., 0. & 
  \multicolumn{2}{|c|}{0.30, 0.55, 0.15} & 0.58, 0.33, 0.09 \\ \hline \hline
  &  & LO & LO & NLO & NLO \\
  & Total cross section [pb] & 0.160 & 0.171 & 0.262 & 0.270 \\ \cline{2-6}
\WmZ  & Fraction of events with jets & 0.286 & \multicolumn{2}{|c|}{0.348} &  
  0.472 \\ \cline{2-6} 
  & Fraction of g,q,\qbar~jets & 1., 0., 0. & 
  \multicolumn{2}{|c|}{0.27, 0.53, 0.20} & 0.56, 0.32, 0.12 \\ \hline \hline
  &  & LO & LO & NLO &  \\
  & Total cross section [pb] & 3.349 & 3.996 & 8.707 & - \\ \cline{2-6}
\Wpg  & Fraction of events with jets & 0.183 & 
  \multicolumn{2}{|c|}{0.438} & - \\ \cline{2-6} 
  & Fraction of g,q,\qbar~jets & 1., 0., 0. & 
  \multicolumn{2}{|c|}{0.18, 0.57, 0.25} & - \\ \hline \hline
  &  & LO & LO & NLO &  \\
  & Total cross section [pb] & 2.226 & 2.681 & 6.418 & - \\ \cline{2-6}
\Wmg  & Fraction of events with jets & 0.180 & 
  \multicolumn{2}{|c|}{0.448} & - \\ \cline{2-6} 
  & Fraction of g,q,\qbar~jets & 1., 0., 0. & 
  \multicolumn{2}{|c|}{0.15, 0.65, 0.20} & - \\ \hline \hline
  &  & LO &  &  & NLO \\
  & Total cross section [pb] & 0.0424 & - & - & 0.0682 \\ \cline{2-6}
ZZ  & Fraction of events with jets & 0.278 & \multicolumn{2}{|c|}{-} &  
  0.365 \\ \cline{2-6} 
  & Fraction of g,q,\qbar~jets & 1., 0., 0. & \multicolumn{2}{|c|}{-} &  
  0.85, 0.11, 0.04\\ \hline \hline
  &  & LO & LO & NLO &  \\
  & Total cross section [pb] & 1.471 & 1.756 & 2.433 & - \\ \cline{2-6}
\Zg  & Fraction of events with jets & 0.190 & \multicolumn{2}{|c|}{0.364} &  
  - \\ \cline{2-6} 
  & Fraction of g,q,\qbar~jets & 1., 0., 0. & 
  \multicolumn{2}{|c|}{0.32, 0.52, 0.16} & -  
  \\ \hline \hline
 \end{tabular}
 \caption{Total cross section, multiplied by the branching fraction
   for decays into electrons and muons, fraction of events with jets
   with \pt$>30$~GeV, and, out of these events, the fractions of 
   events where the jet is due to gluon, quark and anti-quark} 
 \label{tab:xsec}
 \end{center}
\end{table}

For a detailed comparison between the three \MC~programs, large
samples of events have been generated by each program for all
implemented di-boson final states. The CTEQ6M~\cite{CTEQ6} Parton 
Distribution Functions (PDF's) have been used for all \MC~programs,
utilizing the LHAPDF~\cite{LHAPDF} interface. The \Wg~and \Zg~ final 
states have been generated with a 30~GeV lower photon 
\pt~cutoff\footnote{
All transverse momenta in this paper are with respect to the proton 
beam direction.}.
The Z bosons in \ZZ~and \Zg~as produced in \Pythia~can be virtual
and mix with a virtual photon, including interference. 
Therefore, a cut on these
events has been applied, requiring the invariant mass of the generated 
Z decay products to be within $\pm10$~GeV from the Z nominal mass.
All the quoted values and distributions below refer to the events after
these cuts.

The total cross-section values, multiplied by the branching fractions 
for decays into electrons and muons, are listed in
Table~\ref{tab:xsec}. The \Pythia~LO values are lower by 7\% (20\%) 
than the LO values of \BHO~for final states without (with) photon.
The reason for the difference is not clear to us. The electroweak 
parameters
in \Pythia~do not coincide exactly with those of \BHO, but this can
explain a difference of no more than 3\%. The differences between the 
\BHO~and \MCatNLO~cross-section values are smaller, and 
the \BHO~values are lower by 3 - 5\%.
A detailed study to understand the source of these differences is
out of the scope of this paper.  
The K-factors, which are the ratios between the NLO and LO 
cross-section values vary
from 1.3 for \WW~to 2.2 - 2.4 for \Wg. These factors are much
larger than the differences between the various programs, and they 
demonstrate the importance of the NLO corrections.

\begin{figure}[htbp]
\begin{center}
\epsfig{file=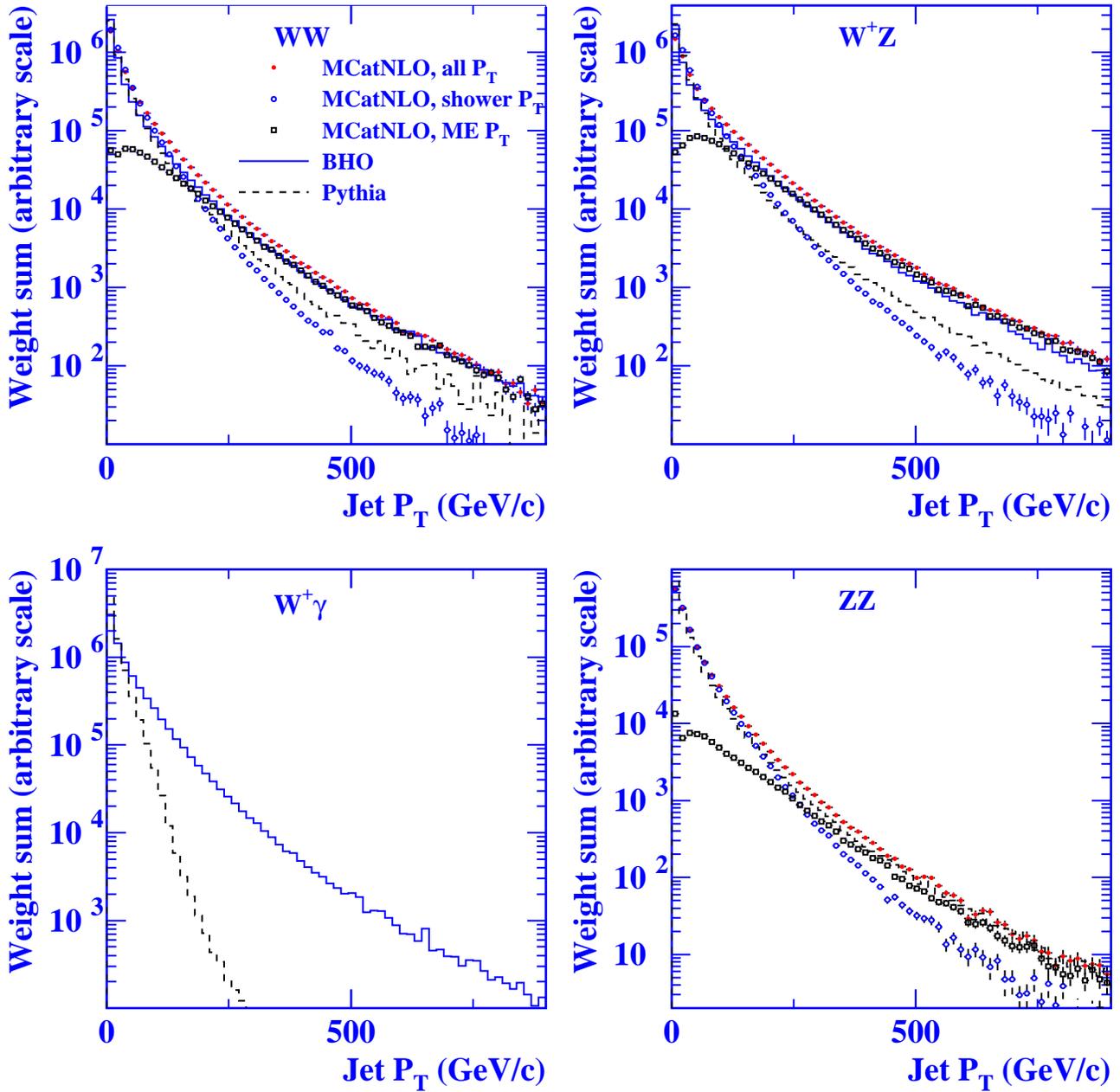,bbllx=30pt,bblly=160pt,bburx=522pt,bbury=645pt,
     width=\textwidth,clip=}
\caption{\sl \pt~distribution of the di-boson system for \WW, \WpZ,
  \Wpg~and ZZ. In each plot, the distributions from the various available 
  \MC~programs are normalized to each other.}
\label{fig:ptjet}
\end{center}
\end{figure}

An interesting kinematic variable to be compared between the
different \MC~programs is the transverse momentum of the di-boson
system, originating from one or more jets which are produced along
the two bosons. This variable is referred below as jet \pt, and events
with jets are defined as events with jet \pt~above 30~GeV. 
Table~\ref{tab:xsec} shows the fraction of events with jets for the 
various \MC~generators. Large differences 
are observed between the different generators. In particular, in
\Pythia, having only SH jets, the fraction of events with jets, as
expected, is always smaller, compared with the other programs. There 
are even larger differences between \BHO~and \MCatNLO. 

For a clearer view, the jet \pt~distributions are plotted in 
Fig.~\ref{fig:ptjet} for \WW, \WpZ, \Wpg~and ZZ events. 
The distributions for \WmZ~(\Wmg~and \Zg) are similar to those for 
\WpZ~(\Wpg). The distributions for \BHO~and \MCatNLO~look similar but 
they are not exactly the same and, as already mentioned, the
fractions of events with jet \pt~above 30 GeV is significantly different. 
The \Pythia~distributions are softer for most cases, differing from 
the other generators, by an amount depending 
strongly on the final state. It is maximal for
final states with photons and almost vanishing for \ZZ. This
observation needs further investigation which is beyond the scope of
this paper. For \MCatNLO~ the 4-momentum of the outgoing ME parton is
available in the generated event record. The rest of the event \pt~is
assigned to the SH partons. The separate contributions of the ME and
SH partons to the jet \pt~are shown in Fig.~\ref{fig:ptjet}. As
expected, the SH \pt~distribution is much softer than the ME one.

Large differences between the \MC~generators exist also in the
relative contributions of the three processes leading to events of 
di-bosons with jets, as listed in Table~\ref{tab:xsec}. The SH partons 
in \Pythia~start from gluons emitted
from the quark lines (Fig.~\ref{fig:feyn}c,d), whereas in \BHO~there
are contributions also from the processes qg$\rightarrow$VVq and 
\qbar g$\rightarrow$VV\qbar, corresponding to Fig.~\ref{fig:feyn}(e-h).
The ratios between these contributions are according to the NLO di-boson 
ME (actually, the LO ME of di-boson + jet). In \MCatNLO~the situation
is more complex. In the NLO process, in anticipation of the following
MC step, the majority of the generated events do not have outgoing
partons. This is demonstrated in the ME \pt~distributions of 
Fig.~\ref{fig:ptjet}, which contain only those events with outgoing
partons and falls below the total \pt~distributions. Adding SH
partons to those events, does not change their classification as
events with outgoing gluon, quark or anti-quark. On the other hand,
all those events generated in the NLO process without outgoing partons, 
and obtain SH partons with \pt~above 30~GeV, are classified as events
with outgoing gluons. This is why the fraction of events with outgoing
gluons is much higher at \MCatNLO~compared to \BHO. However, as
expected, the
ratio between events with outgoing quarks and with outgoing
anti-quarks is approximately the same for these two \MC~programs.

Since the distributions of other kinematic variables are expected to be 
correlated with the jet \pt~distribution and
with the classification of the jet production process, one should not
be surprised to see, for events with jets, some disagreements between those 
distributions. This will be shown in the next section.


\section{Monte Carlo Event Weighting}
 \label{sec:bhowei}

  So far, many \MC~samples of di-boson events have been generated 
by the Tevatron and LHC collaborations using the \Pythia~and
\MCatNLO~programs. These samples 
passed high CPU-consuming detector simulation processes.
As noted in the previous section, the 
\Pythia~events do not include NLO effects. The total cross section and
some differential distributions can be corrected by K-factors, but 
this is not completely satisfactory, since it is not clear whether
those correction factors are not modified by the selection cuts. This
is why more and more samples are produced with \MCatNLO, but as
explained in the previous section, this program also ignores some 
important physics aspects such as spin information for most of the 
di-boson final states and the anomalous TGC's. Fortunately, these 
missing effects exist
in the \BHO~program, but this program cannot be used for event
generation, since it lacks the underlying event and the parton
showering and hadronization processes. 

In order to use the advantages
of the \BHO~program, its ME calculation code has been extracted, so
that it allows calculating for each given generated event a weight value  
which is proportional to the cross section. This can be done
under different conditions. For example, it can include the full decay
of the bosons into leptons using the boson helicity information.
Another possibility is to ignore the boson decay, summing over the
boson helicities. This would imitate the treatment in \MCatNLO.
Taking e.g. an event generated by \MCatNLO, the ratio between the 
weight values calculated with and without boson decay,
can be used to weight that event. The \MCatNLO~sample, after 
this weighting,
should have the correct angular distributions as
expected for events with the spin information. In the same manner, 
weighting by the ratio between the weight values with and without 
an anomalous coupling would introducing the effect of this
coupling to events generated without anomalous couplings, e.g. by 
\Pythia~or \MCatNLO. Any systematic uncertainty related to the ME 
calculation, such as electro-weak parameters, PDF's or $\alpha_s$
can be investigated using a similar weighting.

The ME calculation program handles separately each of the processes 
described in Fig.~\ref{fig:feyn}(a-j), and for each flavor of the
incoming quarks. For events without outgoing partons, only the Born
term is used, in order to avoid negative weights. In any case, for all
the events produced by \Pythia~or \MCatNLO~the di-boson system has some 
transverse momentum value and are then interpreted as events with
one outgoing parton. The momentum of this outgoing parton is the SH
parton momentum, to which one adds the ME parton momentum, if it 
exists, namely, if the event was generated by the NLO step of 
\MCatNLO~with an outgoing parton. There is some ambiguity in the 
calculation of the SH parton momentum. The transverse part is known, 
since it is merely the vector needed to balance the transverse 
momentum vector of the di-boson and ME parton (if present) system. 
The longitudinal part is assumed arbitrarily to vanish in that
system. The identity of the parton is determined as before, namely
it is the identity of the ME parton, if present; otherwise, it is
assumed to be a gluon. The resulting event has all the information
needed for the calculation of its weight under different conditions to
obtain the required weight ratios.

This event weighting method has been tested, as a method to introduce 
the correct spin information, and to introduce anomalous couplings.
This has been done for all di-boson final states. Few examples are
shown below.

\subsection{Spin Information Weighting}

\begin{figure}[htbp]
\begin{center}
\epsfig{file=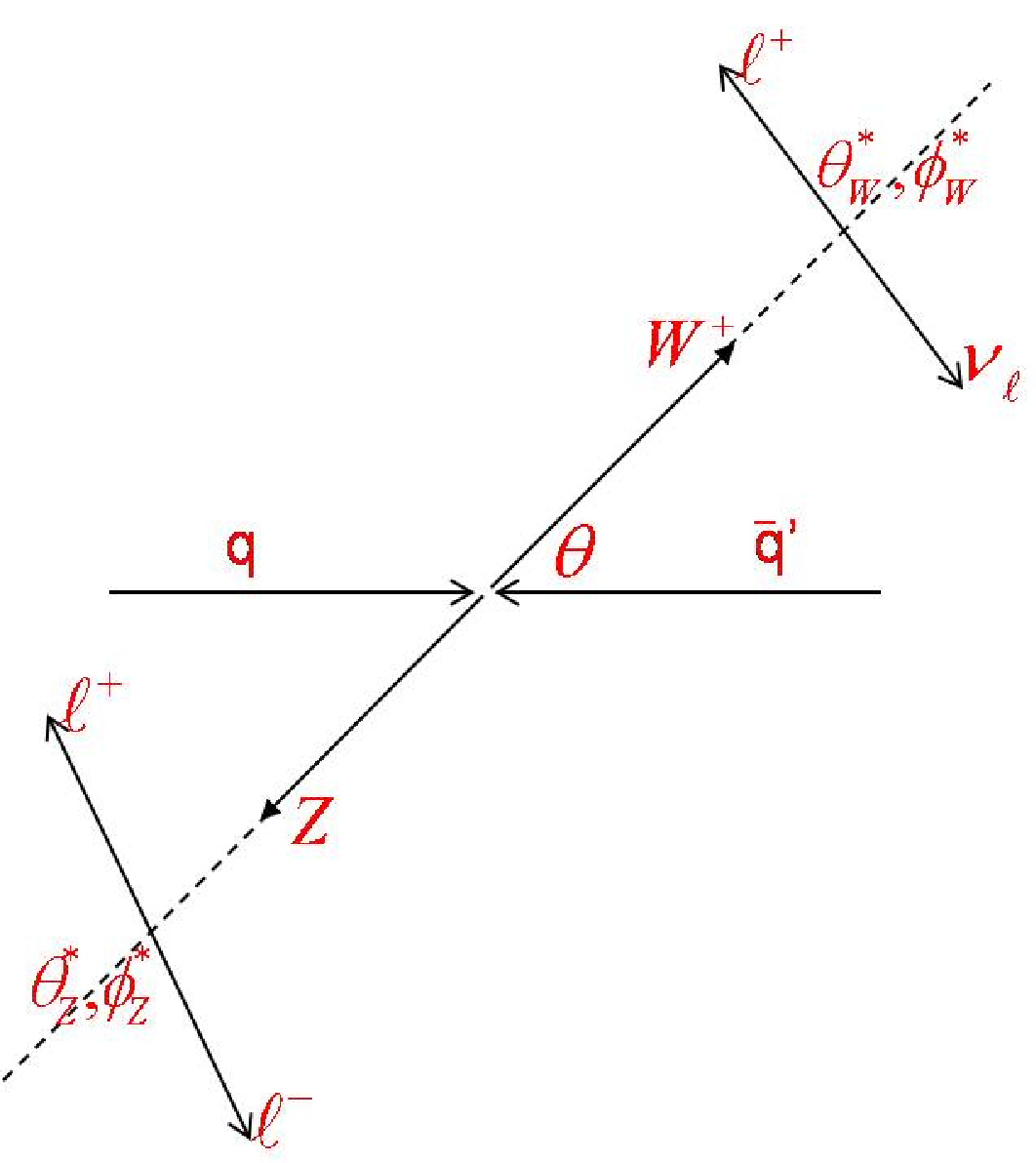,bbllx=0pt,bblly=0pt,bburx=306pt,
 bbury=343pt,width=0.4\textwidth,clip=}
\caption{\sl Production and decay angles of the \WpZ~system}
\label{fig:angdef}
\end{center}
\end{figure}

To test the spin information weighting, it is necessary to look at the
distributions of the production and decay angles. These angles are
defined in Fig.~\ref{fig:angdef}, for \WpZ~events. 

The production 
angle, $\theta$, is the angle between the incoming quark direction and
the outgoing \Wp, at the rest-frame of the \WpZ~system. 
Obviously, Fig.~\ref{fig:angdef} does not describe events with jets,
in particular the case where one of the incoming parton is a gluon.
Even in events without jets the direction of the quark cannot be 
distinguished experimentally from the direction of the anti-quark.
To avoid these difficulties, the direction of the
di-boson system boost in the laboratory frame, ${\bf z_{BB}}$, is used 
instead. This is valid also for events with jets, where it 
does not necessarily coincide with the direction of any 
of the incoming beams. 
This definition relies on the fact that the incoming quark in a proton
beam, which can be a valence quark, is expected, on the average, to be
more energetic than the anti-quark coming always from the sea of the 
opposite proton. Therefore,
for most of the events, the overall boost of the di-boson system is 
expected to be close to the incoming quark direction.

The orbital and azimuthal angles, \Thest,\Phist, of the
charged lepton (anti-lepton) decaying from the Z (\Wp)
are defined in the rest-frame of the decaying boson. The z-axis in
this system is defined as the boson direction in the 
rest-frame of the di-boson system, ${\bf z_B}$.
The y-axis is defined to be orthogonal
to the di-boson production plane, namely, in the direction of 
${\bf z_{BB}}\times{\bf z_B}$, and the x-axis is in the di-boson
production plane, forming together with the two other axes a 
right-handed Cartesian coordinate system. 

\begin{figure}[phtb]
\begin{center}
\epsfig{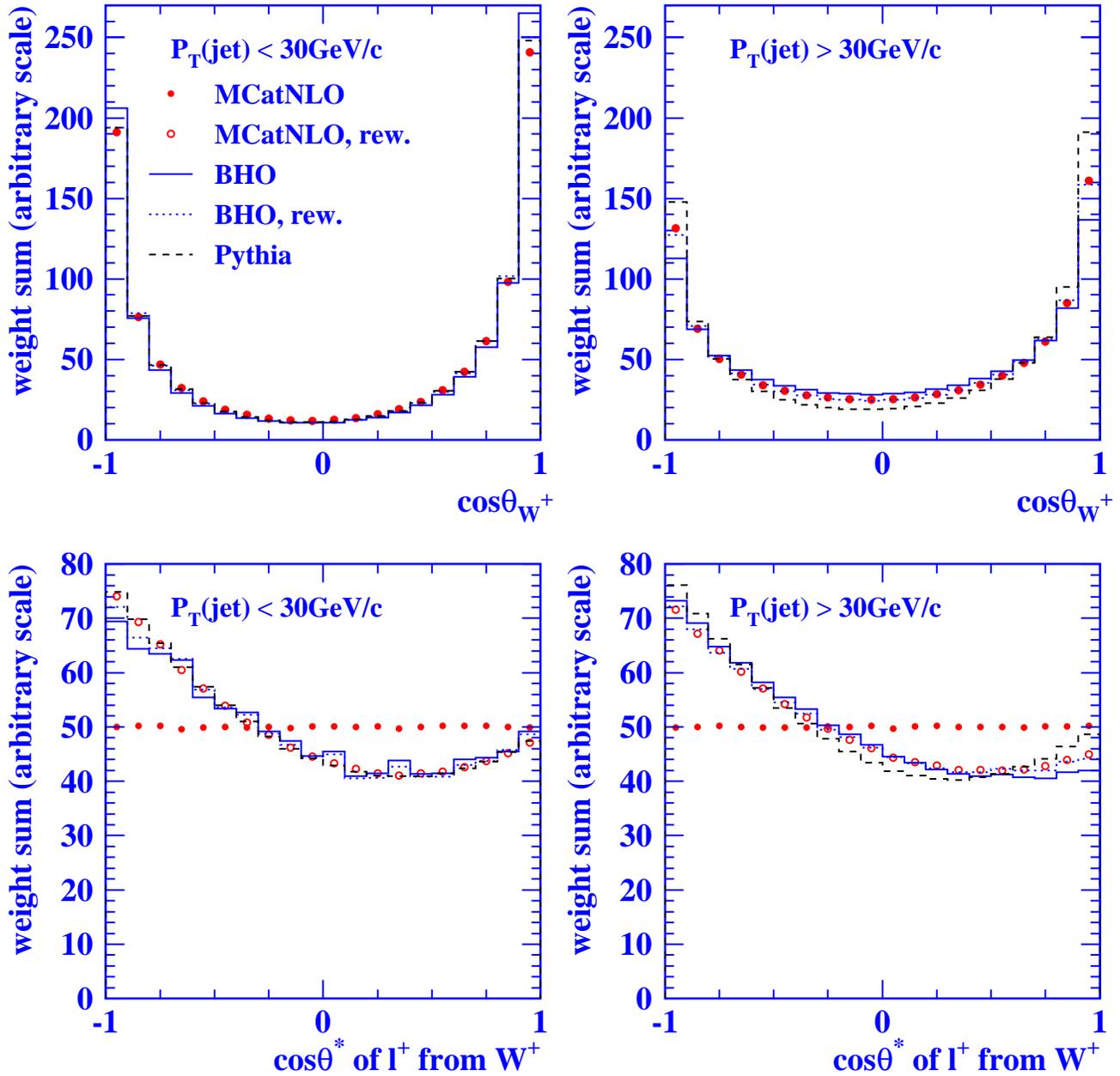}
\caption{\sl Distributions of the W-production and decay angles in 
\WpZ~events. See text for a detailed definition of the angles and 
explanation of the different distributions.}
\label{fig:wpzang1}
\end{center}
\end{figure}

\begin{figure}[phtb]
\begin{center}
\epsfig{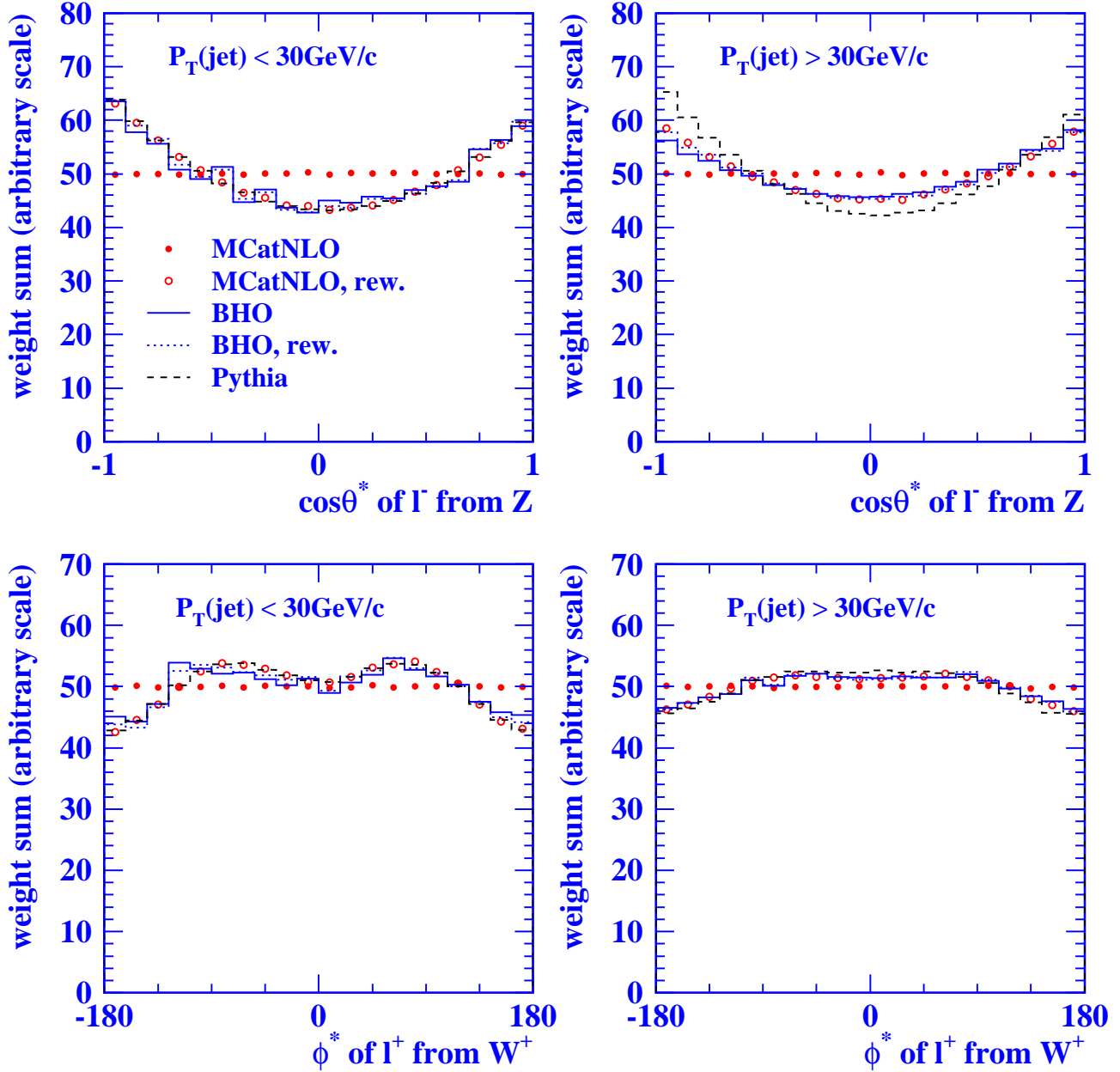}
\caption{\sl Distributions of the orbital Z-decay angle and the 
azimuthal W-decay angle in \WpZ~events. See text for a detailed 
definition of the angles and explanation of the different distributions.}
\label{fig:wpzang2}
\end{center}
\end{figure}

Distributions of the production and some of the decay angles in 
\WpZ~events are shown in Figs.~\ref{fig:wpzang1},\ref{fig:wpzang2},
separately for events without and with jets.
All the distributions are normalized to the same arbitrary total 
sum of weights. 
The distributions for \MCatNLO~before and after the weighting to
correct for the missing spin information are shown by the solid and
open points respectively. For the production angle, the open points
are not visible, since the distributions with and without weighting 
are identical, as expected. Contrarily,
for the decay angles, all the distributions before weighting
are uniform, and after weighting are similar to those from
\Pythia~and \BHO. Some differences do exist, in particular for events
with jets, but they are at the same size as the differences between 
\Pythia~and \BHO. Differences are present also in the production angle
distributions, thus they cannot be entirely attributed to the weighting.
For events without jets, there is a very nice agreement between the
weighted \MCatNLO~and \Pythia, whereas in \BHO~there is a slight 
excess of events with \Wp's produced along the beam axis, and
a small deficiency of \Wp's decaying into a backward charged lepton.
Larger differences are present in the distributions of events with 
jets, in particular between \Pythia~and the other two generators. 
This is not surprising as it was already shown that \Pythia~ does 
not describe properly events with high \pt~jets. It would be 
interesting to see whether also the differences between the 
weighted \MCatNLO~and \BHO~can be related to the differences 
in the jet \pt~distributions and in the fraction of gluon jets 
discussed in the previous section. To check that, the \BHO~events 
have been weighted to give the same number of gluon, quark and 
anti-quark jet events in each \pt~bin. The resulting 
\BHO~distributions after this \pt~weighting (dotted lines in 
Figs.~\ref{fig:wpzang1},\ref{fig:wpzang2}) agree almost perfectly
with \MCatNLO. Even for the low \pt~plots, corresponding to the 
lowest two bins in the \pt~distribution\footnote{
The first \pt~bin in \BHO~includes also events without any jet.
The fraction of these events had to be modified as well, in order
to reach the good agreement with the other generators.},  
the agreement improves,  
in particular for the production angle, where the dotted line 
is not visible as it coincides with the dashed \Pythia~line.

The same comparison of angular distributions has been done for
all other di-boson final states with similar results. For events
with \WW~decaying into leptons, 
the \MCatNLO~generator already contains the spin information. 
Nevertheless, this channel is still included in the code for 
completeness as well as 
for the TGC and other applications. To check the spin 
information part, the inverse of the spin information
weighting has been applied on the events to see if the decay angular 
dependence is removed, and indeed, the decay angle distributions 
become uniform and the production angle distribution is preserved. 
As a further check, the spin information in the W-decays have been
removed by regeneration of the W decay product four-momenta 
corresponding to a uniform decay. These modified events have been
weighted to re-introduce the spin information and the resulting
decay angle distributions have been compared with the original ones.
Again, a very good agreement has been achieved.
As for the \WpZ~case, the agreement with \BHO~is rather good and
improves after the jet \pt~weighting of the \BHO~events.

The \ZZ~final state is not generated by the \BHO~program, but some
part of the \BHO~code covers this channel, without anomalous 
couplings. Consequently, this channel is included in the weighting 
code and has been used to introduce spin information to 
\MCatNLO~events. The agreement of the resulting distributions 
with \Pythia~is excellent for events without jets, and a bit worse
for events with jets, as for the other final states.

The \Wg~and \Zg~channels are not included in \MCatNLO, so the only 
comparison to be done is between \BHO~and \Pythia. The agreement 
between the angular distributions of these two generators is not so
good, unless the \BHO~generator is modified to mimic
\Pythia~as much as possible, namely, using the Born matrix element 
to generate events without jets, and in events with jets, only those 
events with gluon jets are included with jet \pt~weighting.

\subsection{TGC Weighting}

Anomalous TGC's mainly affect events at high c.m. energies of the
hardly interacting partons and large production angles where the 
contribution of the triple gauge vertex $s$-channel diagrams, e.g. in 
Fig.~\ref{fig:feyn}b, is enhanced with respect to the $t$-channel
diagrams (see, e.g. Fig.~\ref{fig:feyn}a). The most convenient
kinematic variable to use, is the transverse momentum of one of the 
outgoing bosons which increases with both c.m. energy and production
angle, and is invariant under the boost of the hardly interacting 
system. For \WW~events, where both W's decay into a charged lepton 
and a neutrino and the W-transverse momentum cannot be reconstructed,
the transverse momentum of the charged lepton is used.

\begin{figure}[phtb]
\begin{center}
\epsfig{file=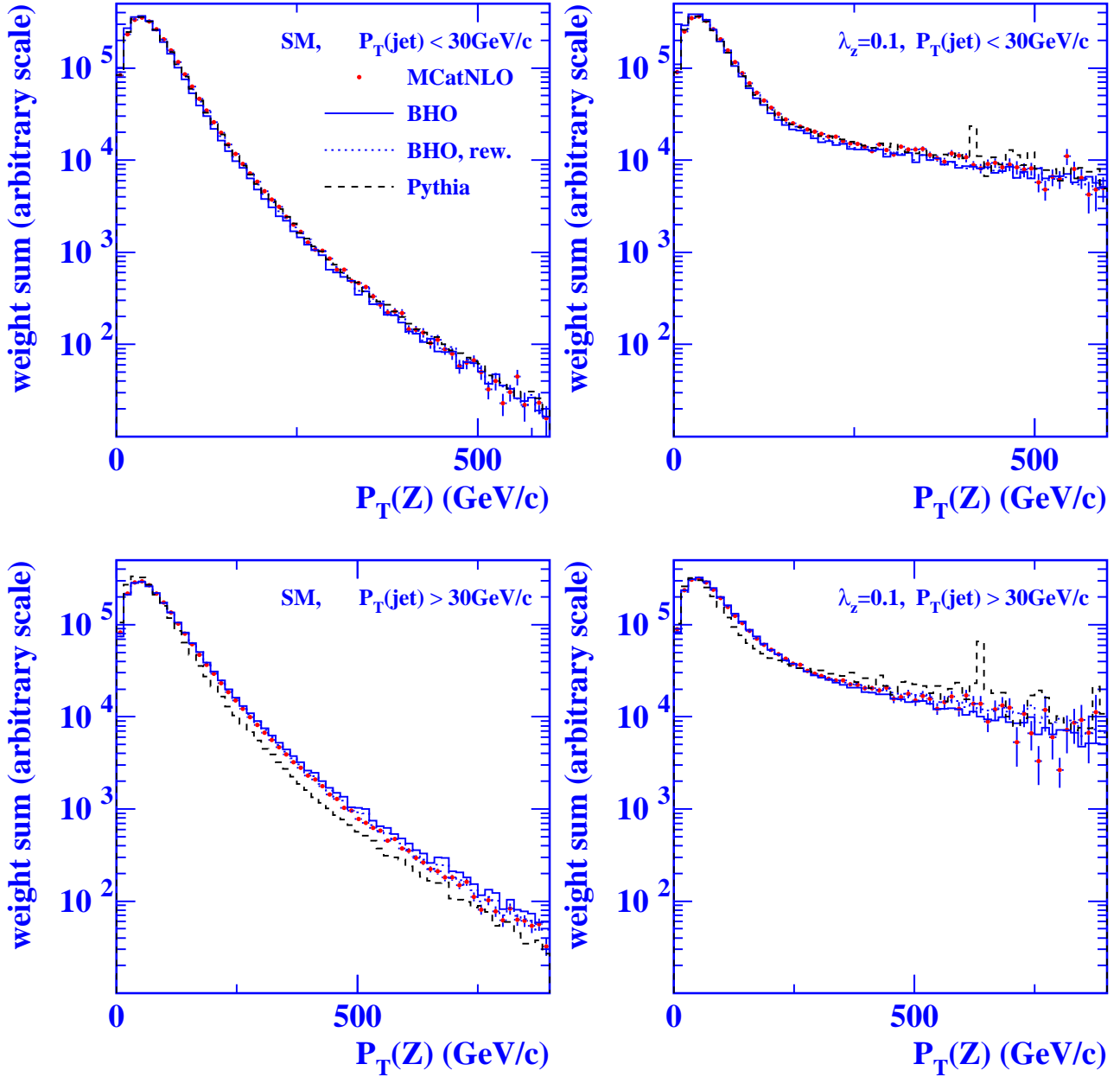,bbllx=25pt,bblly=160pt,bburx=523pt,bbury=696pt,
     width=\textwidth,clip=}
\caption{\sl Transverse momentum distributions of the Z in \WpZ~events
  for the case of the Standard Model and for the case of anomalous
  coupling $\lz=0.1$ separately for events without and with jets}
\label{fig:wpzptz}
\end{center}
\end{figure}

As an example, Fig.~\ref{fig:wpzptz} shows the \pt(Z) distributions for 
\WpZ~events for the case of the Standard Model and for the case of
anomalous coupling $\lz=0.1$. Also here, we distinguish between
events without and with jets. The difference between the distributions
for the Standard Model and for anomalous coupling is remarkable,
already for \pt(Z)~above 150~GeV, and steeply increases with \pt.
The anomalous coupling value used here has been chosen to correspond
to a large effect on the distribution, in order to have a good 
sensitivity to any possible disagreement between the different 
\MC~generators and our calculated weight ratios, which have been 
applied on the \MCatNLO~and the \Pythia~events. 

The \MCatNLO~events have
been weighted to include also the spin information. Their distributions, 
shown by the solid points, have, for events without jets, a nice 
agreement with the \BHO~and \Pythia~distributions (solid and dashed
lines, respectively). This agreement holds for the Standard
Model, as well as for anomalous coupling. For events with jets, the 
agreement is worse, in particular for \Pythia, which, as discussed
above, does not describe well this kind of events. For events
with jets, the discrepancy between \BHO~and \MCatNLO~is less
remarkable, and reduces further when the \BHO~events are weighted 
according to their jet \pt~(dotted line), as described above. 

Similar behavior is seen also for other anomalous couplings, such 
as \dkz~and \dgz. The sensitivity for these couplings is, however,
much smaller, and larger values of the couplings are needed to produce
similar effects on the \pt(Z) distribution.

The same study has been repeated also for the other di-boson channels. 
The \WmZ~channel has been investigated in exactly the same way.
For the \Wg~channels, the Standard Model has been compared with the
anomalous couplings \lgg~and \dkg. For \WW~events, the investigated
TGC's were \lgg, \dkg~and \dgz, assuming the following SU(2)$\times$U(1)
relations between the WW$\gamma$ and WWZ couplings~\cite{LEP2YR,BILENKY}, 
\begin{eqnarray*}
\label{su2u1}
\dkz & = & -\dkg \twsq +  \dgz, \\ 
\lz  & = &  \lgg. 
\end{eqnarray*} 
These relations are motivated by precision measurements on the Z resonance
and lower energy data, and have been used by the LEP 
collaborations~\cite{LEP2}. For the \Zg~channel, the neutral current
anomalous couplings, $h_i^\gamma$ and $h_i^Z$ (i=1,2,3,4) have been 
tested. To avoid unitarity violation, all the anomalous couplings in
the weighting program, following the \BHO~generator, have been
assumed to fall off with increasing invariant mass of the di-boson
system, $M_{\mathrm VV}$ according to the following dipole form factor
relation~\cite{BAURFF},
\beq
\label{dipole}
\alpha(M_{\mathrm VV})=\frac{\alpha(0)}{(1+(M_{\mathrm
    VV}/\Lambda_{\mathrm FF})^2)^n},
\eeq
where $\alpha$ stands for any anomalous TGC, and the form factor scale, 
$\Lambda_{\mathrm FF}$ has been taken as 10~TeV. The power n is 2 for
the WW$\gamma$ and WWZ couplings, whereas for  $h_i^\gamma$ and 
$h_i^Z$ it is 3 (4) for $i$ odd (even). 

\section{Summary and Discussion}

This study of di-boson \MC~event generators shows a nice agreement 
between the differential distributions of various kinematic variables 
for events without jets, although the total cross-section values
differ by few percents. Events with jets, on the other hand, are
generated differently in each generator. Consequently, the jet 
\pt~distributions differ, and even the identities of the partons
forming the jets are not distributed in the same way. These
differences affect the distributions of kinematic variables, including
those which are not directly related with jets. None of the 
\MC~generators is expected to give a precise description of jet 
production, which might be a source of systematic uncertainty in any
analysis which is using \MC~events. This uncertainty can be minimized
in analyses where events with jets are suppressed by the event 
selection cuts. It will also be interesting to have separate studies
of events with jets in the real data, in order to be able to 
choose the most appropriate \MC~generator and tune its free
parameters.

The weighting program described in this paper, when applied to 
 \MCatNLO~and \Pythia~events without jets, seems to give
a sample with the correct kinematic distributions as obtained 
from \BHO. This weighting can be used to
introduce TGC and spin effects wherever needed, as well as 
for investigation of \MC~generator related systematics. 

\begin{figure}[htbp]
\begin{center}
\epsfig{file=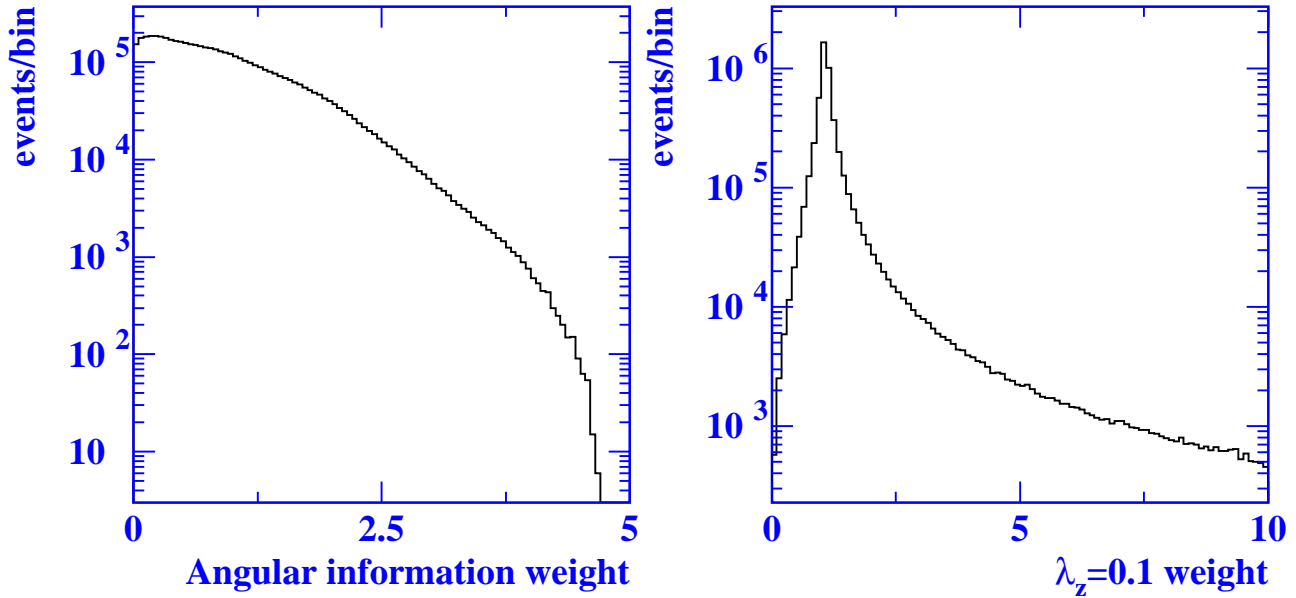,bbllx=25pt,bblly=412pt,bburx=528pt,bbury=646pt,
     width=\textwidth,clip=}
\caption{\sl Distributions of weight ratios used for \WpZ~events to
  introduce spin information and anomalous TGC \lz=0.1}
\label{fig:wpzwei}
\end{center}
\end{figure}

One disadvantage of weighting events is the loss in statistical 
significance, in particular for the case where the distribution
of the weight ratios used for the weighting is broad. 
The weight ratio distributions for \WpZ~events are plotted in 
Fig.~\ref{fig:wpzwei}. The distribution of spin information weight ratios 
is rather narrow without long tails. Applying these weight ratios increases 
the statistical errors of the angular distributions by a factor of 
$\approx1.6$. The distribution of TGC information weight ratios is centered
around 1, but has a long tail at high weight ratio values.  
The events with weight ratio around 1, forming the majority of the sample, 
are those with low \pt(Z) and low sensitivity to anomalous TGC's. 
The events in the tail correspond to high \pt(Z), and contribute most 
of the sensitivity to the TGC analysis. Unfortunately, these very high
weight ratios cause a significant increase in the \MC~statistical errors at
this interesting high \pt~region.  

The finite boson width which is missing in \MCatNLO~events cannot 
be introduced by event weighting. This problem, however, is not
expected to have a significant effect, unless the di-bosons are
produced close to their kinematic threshold. The nice agreement
for events without jets between the distributions of \Pythia~and 
\MCatNLO~demonstrates that the boson width in \Pythia~does not 
play an important r\^{o}le.      
  

\section*{Appendix: A Short Manual of the Weighting 
Package}
The weighting package is available on
{\tt http://atlas2.tau.ac.il/bella/bhowei.html} in two
FORTRAN files. The first one contains the main subroutine, 
{\tt bhowei} and other subroutines called by the main one. 
All these routines have been rewritten following the original 
\BHO~code. The second file contains other subroutines which were
taken from the original \BHO~code without modification. In addition 
to these two files, the user has to link the LHAPDF library which can 
be downloaded from {\tt http://projects.hepforge.org/lhapdf/}.
The calling sequence is,

{\tt call bhowei(mspin,xsec,ifail)}

where,

\begin{tabular}{lll}
{\tt mspin} & (input, integer) & 1 if spin information is needed, otherwise,
0 \\
{\tt xsec} & (output, real) & calculated weight, proportional to the
cross section \\
{\tt ifail} & (output, integer) & different from 0 in case of a problem
\end{tabular} 

All the other information needed to run the program is inserted via
common blocks.

The event data is inserted into the following common block,

{\tt common/cproc/id(7),p(0:3,7)}

where {\tt id} is an integer array of the PDG Monte Carlo particle 
codes~\cite{PDG}
of the seven particles involved in the di-boson production. The first
two are the incoming partons, the next four are the decay products of
the two bosons and the last one is the outgoing parton. In case of
a photon, its code, 22, is inserted in the position of the first decay
product and the position of the second is filled with 0. 
Similarly, {\tt id(7)}=0 in case of no outgoing parton. The real array
{\tt p} must contain the four-vectors of these particles at the same 
order as in {\tt id}.    

\begin{table}[thbp]
 \begin{center}
 \begin{tabular}{|l|c|c|l|} \hline
 Variable & Type & Default & Description \\ \hline
 {\tt is} & integer & 1 & 1 for pp, -1 for p$\bar{\mathrm p}$ \\
 {\tt npdf} & integer & 10000 & LHAGLUE number of PDF set \\
 {\tt loop} & integer & 2 & number of loops in $\alpha_s$ calculation \\
 {\tt cq} & real & 1. & coefficient used to define $q^2$ (scale of 
                $\alpha_s$ and PDF) \\
 {\tt iscale} & integer & 2 & $q^2$ definition flag,  \hspace{2mm} =1
        for $q^2=${\tt cq}$\cdot$Mass(diboson+jet)$^2$, \\
  &  &  & \hspace{2mm} =2 for $q^2$={\tt cq}$\cdot$Mass(diboson)$^2$,
     \hspace{2mm} =3 for $q^2$={\tt cq}$\cdot$\pt(jet)$^2$,  \\
  &  &  & \hspace{2mm} =4 for $q^2$={\tt cq}$\cdot$Mass(W)$^2$,
     \hspace{11mm} =5 for $q^2$={\tt cq}$\cdot${\tt mscale}$^2$  \\
 {\tt mscale} & real & 100. & used to define $q^2$ for {\tt iscale}=5
 \\ 
 {\tt ecm} & real & 14000. & c.m. energy of the interacting protons \\
 {\tt xlambda4} & real &    & $\Lambda_{\mathrm QCD}$ as calculated by
 LHAPDF \\ 
 {\tt lambda\_scale} & real & 10000. & TGC form factor scale, 
  $\Lambda_{FF}$ \\
 {\tt mw} & real & 80.40 & W mass \\
 {\tt mz} & real & 91.187 & Z mass \\
 {\tt mt} & real & 175 & t-quark mass \\
 {\tt gw} & real & 2.12 & W width \\
 {\tt gz} & real & 2.487 & Z width \\
 {\tt alfs} & real & 0.116 & $\alpha_s(m_Z^2)$ \\
 {\tt alfem} & real & 1/128. & $\alpha_{EM}(m_Z^2)$ \\
 {\tt coscab2} & real & 0.95 & $\cos^2\theta_c$, $\theta_c$ is the
       Cabibbo angle \\ \hline 
 \end{tabular}
 \caption{Parameters in {\tt common/const/}. 
          All energy values are in GeV.}
 \label{tab:const}
 \end{center} 
\end{table}

Constant parameters needed for the calculation are introduced in,
 
{\tt common/const/is,npdf,loop,iscale,ecm,cq,mscale,xlambda4,} \\
\hspace*{27mm}{\tt lambda\_scale,mw,mz,mtop,gw,gz,alfs,alfem,coscab2}

as explained in Table~\ref{tab:const}. Any calculation in the program
using these constants is done for each call to {\tt bhowei} and not
just at the beginning of the run. Therefore, they can be modified 
before calling {\tt bhowei}, and this can be done several times for 
each events with different parameter values in order to investigate 
their related systematics. 

Anomalous couplings can be introduced via the following common blocks,

{\tt common/ctgc/dg1z,dkz,lamz,dkg,lamg} \\
{\tt common/ntgc/hz(4),hg(4)}

where {\tt dg1z, dkz, lamz, dkg, lamg} are \dgz, \dkz, \lz, \dkg, 
\lgg, and {\tt hz(4), hg(4)} are  $h_i^Z$ and $h_i^\gamma$
($i$=1,2,3,4) respectively.
  

\section*{Acknowledgments}

I wish to thank U. Baur for kindly providing me with the BHO code.
Many thanks to my colleagues in the ATLAS group at Tel Aviv University
for their support, and in particular to E. Etzion for reading the 
manuscript. This research was supported by the Israel Science 
Foundation.  

\bibliography{dibosmc}

\end{document}